\documentstyle[amssymb,preprint,prl,aps]{revtex}

\begin{document}
\draft
\title{Study on two coupled relativistic superfluids with spontaneous symmetry
breaking}
\author{Sun Zhang}
\address{Department of Physics, Nanjing University, Nanjing 210093, China}
\maketitle

\begin{abstract}
In this Letter, we have studied two coupled relativistic superfluids with
spontaneous U(1) symmetry breaking, using Poisson bracket technique. After
constructing the commutators between thermo-quantities and field quantities,
the equations of motion are obtained. These equations describe the system in
the frame of two-constituent superfluid theory and provide a clear picture
relating the symmetry breaking and the physical quantities explicitly. The
interference effect is discussed and the Josephson-type equation is also
given. Furthermore, the dissipation effect and entropy production mechanism
are presented and the dissipative coefficients are given explicitly in the
first-order theory framework.
\end{abstract}


In 1941, Landau proposed his famous two-constituent theory on superfluid $^4$%
He \cite{1,2}. The central idea of the theory is that at temperature below
the critical point, helium behaves as if it were a mixture of two different
liquids. One of these is a superfluid, and moves with zero viscosity along a
solid surface, while the other is a normal viscous liquid. No momentum is
transferred from one to the other. The two-constituent superfluid theory was
studied and generalized by Khalatnikov \cite{3} brilliantly in the
non-relativistic frame.

The first approach to the relativistic superfluid theory was provided by
Israel \cite{4} and Dixon \cite{5,6} concerning the perfect fluids based on
the idea of two-constituents originated by Landau. The medium is usually
called relativistic in two senses: if it has a relativistic equation of
state or when it flows at a relativistic velocity. While these conditions
can be both satisfied in the neutron stars whose interior construction is
composed of a superfluid nuclear matter \cite{7}. The method proposed by
Israel and Dixon is useful for some calculations, but arising a problem.

Generally speaking, the coupled two-constituents are not perfect fluids
strictly and the superfluid component cannot be regarded as a completely
independent fluid \cite{8}. That is, the coupling effects lead the deviation
from ideality \cite{9}. It is meaningful to take into account the deviation
from the perfect fluids. Khalatnikov and Lebedev \cite{8} and Carter and
Khalatnikov \cite{10} suggested two equivalent approaches to include the
interaction between the superfluid and normal fluid.

Furthermore, Son \cite{11} presented another approach to the relativistic
superfluid with U(1) symmetry breaking based on the Poisson bracket
technique \cite{12}. Such a description is also arisen from the
two-constituent theory, but it is more clearer to show the relation between
superfluid and the symmetry breaking. After constructing some basic
relations of the fundamental quantities by Poisson brackets, the dynamic
equations are obtained directly, which provide the explicit meaning about
the symmetry breaking.

It is possible to generalize the discussion presented in Ref. \cite{11} in
several directions. The most important one is that we should construct a
clear frame to discuss the effects between two coupled relativistic
superfluids including the coherence effect and the dissipation effect. For
the symmetry spontaneous breaking and the phase coherence are assumed from
the very beginning, such an approach could be looked as a natural framework
for investigating the inter-relationship of two coupled relativistic
superfluids in the hydrodynamic limit.

To study the two coupled relativistic superfluids 1 and 2 with U(1) symmetry
breaking, the order parameter is of the form $\langle \psi _i\rangle =\left|
\langle \psi _i\rangle \right| e^{i\phi _i}$ ($i=1,2$), the density of the
U(1) charge $\rho _i=-i(\psi _i^{*}\partial _0\psi _i-\partial _0\psi
_i^{*}\psi _i)$ and the phase $\phi _i$ are introduced \cite{11}.

In order to establish the dynamic equations, the Poisson brackets should be
written down to give the commutative relations between dynamic variables.
There are two kinds of Poisson brackets: the fluid (thermo-) ones and the
field ones.

The fluid ones are \cite{11,12} 
\begin{equation}
\lbrack T^{0i}({\bf x}),A({\bf y})]=A({\bf x})\partial _i\delta ({\bf x}-%
{\bf y}),\qquad A=\rho _1,\rho _2,s  \label{108}
\end{equation}
\begin{equation}
\lbrack T^{0i}({\bf x}),B({\bf y})]=-\partial _iB({\bf x})\delta ({\bf x}-%
{\bf y}),\qquad B=\phi _1,\phi _2  \label{109}
\end{equation}
\begin{equation}
\lbrack T^{0i}({\bf x}),T^{0j}({\bf y})]=\left( T^{0j}({\bf x})\frac \partial
{\partial x^i}-T^{0i}({\bf y})\frac \partial {\partial y^j}\right) \delta (%
{\bf x}-{\bf y}),  \label{110}
\end{equation}
where $T^{0i}$ is the momentum density in the 4-dimensional energy-momentum
tensor, $s$ the entropy density. $\rho _{1,2}$ and $\varphi _{1,2}$ are just
the density of charge and the phase of the order parameter respectively.

The Poisson bracket for the field part includes the commutator of $\rho _i$
with $\phi _j$ (up to an adjustable factor) 
\begin{equation}
\lbrack \rho _i({\bf x}),\phi _j({\bf y})]=-\delta ({\bf x}-{\bf y})\delta
_{ij}.\qquad i,j=1,2  \label{111}
\end{equation}

To get the equation of motion, we should construct the most general
Hamiltonian dependent on the fluid quantities and the field quantities 
\begin{equation}
H=\int d{\bf r}T^{00}(s,\rho _1,\rho _2,T^{0i},\partial _i\phi _1,\partial
_i\phi _2),  \label{113}
\end{equation}
where $T^{00}$ is a functional of $\partial _i\phi _{1,2}$ but not $\phi
_{1,2}$ itself due to the U(1) invariance \cite{13,14}. The form of the $%
T^{00}$ cannot be identified at this moment, it could be calculated from the
microscopic theory.

Then the equation of state, describing the relation of the pressure $p$, the
temperature $T$, the chemical potentials $\mu _{1,2}$, and the phases $\phi
_{1,2}$, could be written in the differential form as 
\begin{equation}
dp=sdT+\rho _1d\mu _1+\rho _2d\mu _2+V_{11}^2d\left( \frac 12(\partial _\mu
\phi _1)^2\right) +V_{22}^2d\left( \frac 12(\partial _\mu \phi _2)^2\right)
-(V_{12}^2+V_{21}^2)d\left( \frac 12\partial _\mu \phi _1\partial ^\mu \phi
_2\right) ,  \label{114}
\end{equation}
where, $V_{11}^2$ and $V_{22}^2$ are the factors characterizing the strength
of the internal energy for superfluids 1 and 2 respectively, while $V_{12}^2$
and $V_{21}^2$ are the coupling constants between two superfluids due to
interference and $V_{12}^2=V_{21}^2$. In Eq. (\ref{114}), the first three
terms are fluid terms, they describe the normal fluids part, while the
gradients of the phases in the left terms are specific to the superfluids
and have the significance of the superfluid velocities.

Based on the Poisson brackets (\ref{108})---(\ref{111}) and the
thermodynamic relations, we obtain the dynamic equations 
\begin{equation}
\partial _\mu \left( u^\mu \rho _1-V_{11}^2\partial ^\mu \phi _1+\frac 12%
(V_{12}^2+V_{21}^2)\partial ^\mu \phi _2\right) =0,  \label{140}
\end{equation}
\begin{equation}
\partial _\mu \left( u^\mu \rho _2-V_{22}^2\partial ^\mu \phi _2+\frac 12%
(V_{12}^2+V_{21}^2)\partial ^\mu \phi _1\right) =0,  \label{141}
\end{equation}
\begin{equation}
u^\mu \partial _\mu \phi _1+\mu _1=0,  \label{142}
\end{equation}
\begin{equation}
u^\mu \partial _\mu \phi _2+\mu _2=0.  \label{143}
\end{equation}
It is clear that Eqs. (\ref{140}) and (\ref{141}) are the conservation laws
of the U(1) charge in the Lorentz covariance form. Within the bracket of Eq.
(\ref{140}) [or Eq. (\ref{141})], the first term presents the normal current
and the 4-dimensional velocity $u^\mu $ is the velocity of the normal
component; while the second term gives the superfluid current and the
gradient of the phase is clearly the velocity of the superfluid component,
and the two-constituent description is constructed explicitly. The third
term is the interference current, which illuminates the phase coherent
influence of the superfluids from each other. Then the U(1) charge is a sum
of the normal current, the superfluid current, and the interference current
due to the coupling to the other superfluid, and Eqs. (\ref{140}) and (\ref
{141}) give the conservation relations of the U(1) charge.

On the other hand, Eq. (\ref{142}) [or Eq. (\ref{143})] gives the time
dependence of the condensate phase on the chemical potential. The chemical
potential in these equations is just the change in energy density per change
in the charge density, and also the minimum energy required to add a paticle
to the system. The chemical potential also has the meaning of the background
field due to the U(1) gauge transformation, and coincides the scalar
potential $V$ in the superconductivity. If we suppose that the two
superfluids are kept at uniform chemical potentials, it is direct to obtain
the time dependent relation of the difference of the condensate phases
according to Eqs. (\ref{142}) and (\ref{143}) 
\begin{equation}
u^\mu \partial _\mu (\phi _1-\phi _2)+(\mu _1-\mu _2)=0.  \label{144}
\end{equation}
It is just the relativistic version of the Josephson equation, in BEC
coherence effect \cite{15} and superconductivity \cite{16}.

We can also construct the energy-momentum tensor $T^{\mu \nu }$ as a sum of
the thermo (fluid) part and field (superfluid) part as follows

\begin{equation}
T^{\mu \nu }=(\epsilon +p)u^\mu u^\nu -pg^{\mu \nu }+V_{11}^2\partial ^\mu
\phi _1\partial ^\nu \phi _1+V_{22}^2\partial ^\mu \phi _2\partial ^\nu \phi
_2-V_{12}^2\partial ^\mu \phi _1\partial ^\nu \phi _2-V_{21}^2\partial ^\mu
\phi _2\partial ^\nu \phi _1,  \label{12}
\end{equation}
which satisfies the conservation law 
\begin{equation}
\partial _\mu T^{\mu \nu }=0,  \label{120.5}
\end{equation}
where the fluid part \cite{2} 
\begin{equation}
T_{\text{fluid}}^{\mu \nu }=(\epsilon +p)u^\mu u^\nu -pg^{\mu \nu },
\label{121}
\end{equation}
describes the normal relativistic fluid outside the superfluid condensation
phase and the metric tensor is 
\begin{equation}
g^{\mu \nu }=\text{diag}(1,-1,-1,-1),  \label{122}
\end{equation}
in accordance with Landau and Lifshitz, and $u^\mu u_\mu =1$. While the
field part describes the coherent motion and the interference effect of two
superfluids.

Next, we would like to explore the dissipation effect and the entropy
production mechanism of two coupled relativistic superfluids system to
provide a full understanding of the behaviors described here, following
Landau and Lifshitz \cite{2} and Weinberg \cite{17}.

Generally speaking, the dissipative processes are important, when space-time
gradients of hydrodynamic quantities in the system become large relative to
its relaxation scales. We would analyze the dynamics of small departures of
the system from equilibrium states and manage to give the dissipative
equations as linear responses to the deviations from equilibrium states and
two famous dissipative theories by Eckart \cite{18} and by Landau and
Lifshitz \cite{2} would be the special cases in the theory presented here.

When we study the imperfect system of two coupled superfluids, the
dissipative effects should be taken into account by modifying the
energy-momentum tensor $T^{\mu \nu }$ by a term $\Delta T^{\mu \nu }$, and
could be written in matrix form as 
\begin{equation}
T^{\mu \nu }+\Delta T^{\mu \nu }=(\epsilon +p)u^\mu u^\nu -pg^{\mu \nu
}+\left( 
\begin{array}{ll}
\partial ^\mu \phi _1, & \partial ^\mu \phi _2
\end{array}
\right) \left( 
\begin{array}{ll}
~~V_{11}^2~ & -V_{12}^2 \\ 
-V_{21}^2~ & ~~V_{22}^2
\end{array}
\right) \left( 
\begin{array}{l}
\partial ^\nu \phi _1 \\ 
\partial ^\nu \phi _2
\end{array}
\right) +\Delta T^{\mu \nu },  \label{145}
\end{equation}
satisfying the conservation equation as 
\begin{equation}
\partial _\mu (T^{\mu \nu }+\Delta T^{\mu \nu })=0.  \label{146}
\end{equation}
The other dissipative effects for the charge and the phase could be
considered as 
\begin{equation}
\partial _\mu \left( 
\begin{array}{l}
u^\mu \rho _1+\Delta \rho _1^\mu \\ 
u^\mu \rho _2+\Delta \rho _2^\mu
\end{array}
\right) =\partial _\mu \left[ \left( 
\begin{array}{ll}
~~V_{11}^2~ & -V_{12}^2 \\ 
-V_{21}^2~ & ~~V_{22}^2
\end{array}
\right) \left( 
\begin{array}{l}
\partial ^\mu \phi _1 \\ 
\partial ^\mu \phi _2
\end{array}
\right) \right] ,  \label{147}
\end{equation}
and 
\begin{equation}
u^\mu \partial _\mu \left( 
\begin{array}{l}
\phi _1 \\ 
\phi _2
\end{array}
\right) =-\left( 
\begin{array}{l}
\mu _1+\Delta \phi _1 \\ 
\mu _2+\Delta \phi _2
\end{array}
\right) .  \label{148}
\end{equation}
The dissipative terms should satisfy the following constraints 
\begin{equation}
u_\mu u_\nu \Delta T^{\mu \nu }=0,  \label{149}
\end{equation}
\begin{equation}
u_\mu \Delta \rho _{1,2}^\mu =0.  \label{150}
\end{equation}
For comparison, the constraint used by Landau and Lifshitz is $u_\nu \Delta
T^{\mu \nu }=0$ and that of Eckart is $\Delta \rho ^\mu =0$.

Then using the equations of motion and the thermodynamic relations, we can
get the entropy production equation arisen by the dissipative terms, 
\begin{eqnarray}
\partial _\mu S^\mu &=&-\partial _\mu \left( \frac{\mu _1}T~,~\frac{\mu _2}T%
\right) \left( 
\begin{array}{l}
\Delta \rho _1^\mu \\ 
\Delta \rho _2^\mu
\end{array}
\right)  \nonumber \\
&&+\left( \frac{\Delta \phi _1}T~,~\frac{\Delta \phi _2}T\right) \partial
_\mu \left[ \left( 
\begin{array}{ll}
~~V_{11}^2~ & -V_{12}^2 \\ 
-V_{21}^2~ & ~~V_{22}^2
\end{array}
\right) \left( 
\begin{array}{l}
\partial ^\mu \phi _1 \\ 
\partial ^\mu \phi _2
\end{array}
\right) \right]  \nonumber \\
&&+\frac 1T\Delta T^{\mu \nu }\partial _\mu u_\nu -\frac 1{T^2}u_\mu \Delta
T^{\mu \nu }\partial _\nu T.  \label{151}
\end{eqnarray}
Where the entropy $S^\mu $ density is defined as 
\begin{equation}
S^\mu =u^\mu s-\left( \frac{\mu _1}T~,~\frac{\mu _2}T\right) \left( 
\begin{array}{l}
\Delta \rho _1^\mu \\ 
\Delta \rho _2^\mu
\end{array}
\right) +\frac 1Tu_\nu \Delta T^{\mu \nu }.  \label{152}
\end{equation}
By means of the projection operator ($g^{\mu \nu }-u^\mu u^\nu $), $\Delta
T^{\mu \nu }$ can be expressed as 
\begin{eqnarray}
\Delta T^{\mu \nu } &=&\left( (g^{\mu \alpha }-u^\mu u^\alpha )u^\nu
+(g^{\nu \alpha }-u^\nu u^\alpha )u^\mu \right) q_\alpha  \nonumber \\
&&+(g^{\mu \alpha }-u^\mu u^\alpha )(g^{\nu \beta }-u^\nu u^\beta )\tau
_{\alpha \beta }+(g^{\mu \nu }-u^\mu u^\nu )\tau .  \label{153}
\end{eqnarray}
Defining the quantities $\Delta N_{1,2}$ as 
\begin{equation}
\left( 
\begin{array}{l}
\Delta N_1 \\ 
\Delta N_2
\end{array}
\right) =\left( 
\begin{array}{ll}
\sigma _{11} & \sigma _{12} \\ 
\sigma _{21} & \sigma _{22}
\end{array}
\right) \left( 
\begin{array}{l}
\frac{\mu _1}T \\ 
\frac{\mu _2}T
\end{array}
\right) ,  \label{154}
\end{equation}
where $\sigma _{11}$ and $\sigma _{22}$ are the charge diffusion
coefficients for superfluids 1 and 2 respectively, while $\sigma
_{12}=\sigma _{21}$ are the mutual diffusion coefficients between two
superfluids and are supposed to have only nonnegative eigenvalues. Then the
charge dissipative equations are 
\begin{eqnarray}
\left( 
\begin{array}{l}
\Delta \rho _1^\mu \\ 
\Delta \rho _2^\mu
\end{array}
\right) &=&(g^{\mu \nu }-u^\mu u^\nu )\partial _\nu \left[ \left( 
\begin{array}{ll}
\sigma _{11} & ~\sigma _{12} \\ 
\sigma _{21} & ~\sigma _{22}
\end{array}
\right) \left( 
\begin{array}{l}
\frac{\mu _1}T \\ 
\frac{\mu _2}T
\end{array}
\right) \right]  \nonumber \\
&=&(g^{\mu \nu }-u^\mu u^\nu )\partial _\nu \left( 
\begin{array}{l}
\Delta N_1 \\ 
\Delta N_2
\end{array}
\right) .  \label{155}
\end{eqnarray}
Then the first term in Eq. (\ref{151}) could be written as 
\begin{eqnarray}
-\partial _\mu \left( \frac{\mu _1}T~,~\frac{\mu _2}T\right) \left( 
\begin{array}{l}
\Delta \rho _1^\mu \\ 
\Delta \rho _2^\mu
\end{array}
\right) &=&-(g^{\mu \nu }-u^\mu u^\nu )\partial _\mu \left( \frac{\mu _1}T~,~%
\frac{\mu _2}T\right) \partial _\nu \left( 
\begin{array}{l}
\Delta N_1 \\ 
\Delta N_2
\end{array}
\right)  \nonumber \\
&=&-(g^{\mu \nu }-u^\mu u^\nu )\partial _\mu \left( \Delta N_1~,~\Delta
N_2\right) \left( 
\begin{array}{ll}
\sigma _{11} & ~\sigma _{12} \\ 
\sigma _{21} & ~\sigma _{22}
\end{array}
\right) ^{-1}\partial _\nu \left( 
\begin{array}{l}
\Delta N_1 \\ 
\Delta N_2
\end{array}
\right) ,  \label{156}
\end{eqnarray}
where the positivity of the inverse matrix of $\sigma $ is assumed. Defining
the quantities $A_{1,2}^\mu $ as 
\begin{equation}
\left( 
\begin{array}{l}
A_1^\mu \\ 
A_2^\mu
\end{array}
\right) =\left( 
\begin{array}{ll}
~~V_{11}^2~ & -V_{12}^2 \\ 
-V_{21}^2~ & ~~V_{22}^2
\end{array}
\right) \left( 
\begin{array}{l}
\partial ^\mu \phi _1 \\ 
\partial ^\mu \phi _2
\end{array}
\right) ,  \label{157}
\end{equation}
and $\Delta M_{1,2}^\mu $ as 
\begin{equation}
\left( 
\begin{array}{l}
\Delta M_1^\mu \\ 
\Delta M_2^\mu
\end{array}
\right) =\left( 
\begin{array}{ll}
\zeta _{11} & ~\zeta _{12} \\ 
\zeta _{21} & ~\zeta _{22}
\end{array}
\right) \left( 
\begin{array}{l}
A_1^\mu \\ 
A_2^\mu
\end{array}
\right) ,  \label{158}
\end{equation}
where $\zeta _{11}$ and $\zeta _{22}$ are the bulk viscosity coefficients
for superfluids 1 and 2 respectively, while $\zeta _{12}=\zeta _{21}$ are
the mutual viscosity coefficients between two superfluids and have only
nonnegative eigenvalues supposedly.

Then the second term in Eq. (\ref{151}) could be written as 
\begin{eqnarray}
&&\left( \frac{\Delta \phi _1}T~,~\frac{\Delta \phi _2}T\right) \partial
_\mu \left[ \left( 
\begin{array}{ll}
~~V_{11}^2~ & -V_{12}^2 \\ 
-V_{21}^2~ & ~~V_{22}^2
\end{array}
\right) \left( 
\begin{array}{l}
\partial ^\mu \phi _1 \\ 
\partial ^\mu \phi _2
\end{array}
\right) \right]   \nonumber \\
&=&\left( \frac{\Delta \phi _1}T~,~\frac{\Delta \phi _2}T\right) \partial
_\mu \left( 
\begin{array}{l}
A_1^\mu  \\ 
A_2^\mu 
\end{array}
\right) =\frac 1T\partial _\mu \left( \Delta M_1^\mu ~,~\Delta M_2^\mu
\right) \left( 
\begin{array}{ll}
\zeta _{11} & ~\zeta _{12} \\ 
\zeta _{21} & ~\zeta _{22}
\end{array}
\right) ^{-1}\partial _\nu \left( 
\begin{array}{l}
\Delta M_1^\nu  \\ 
\Delta M_2^\nu 
\end{array}
\right)   \nonumber \\
&&+\frac 1T\partial _\mu u^\mu (\zeta _1~,~\zeta _2)\partial _\nu \left( 
\begin{array}{l}
A_1^\nu  \\ 
A_2^\nu 
\end{array}
\right) ,  \label{159}
\end{eqnarray}
if we present the dissipative equation $\Delta \phi _{1,2}$ as 
\begin{equation}
\left( 
\begin{array}{l}
\Delta \phi _1 \\ 
\Delta \phi _2
\end{array}
\right) =\left( 
\begin{array}{ll}
\zeta _{11} & ~\zeta _{12} \\ 
\zeta _{21} & ~\zeta _{22}
\end{array}
\right) \partial _\mu \left( 
\begin{array}{l}
A_1^\mu  \\ 
A_2^\mu 
\end{array}
\right) +\partial _\mu u^\mu \left( 
\begin{array}{l}
\zeta _1 \\ 
\zeta _2
\end{array}
\right) ,  \label{160}
\end{equation}
where the viscosity coefficients $\zeta _1$ and $\zeta _2$ are those related
to the normal velocity of the systems respectively. The dissipative
components in $\Delta T^{\mu \nu }$ of Eq. (\ref{153}) could be shown as 
\begin{equation}
q_\alpha =\kappa T\left( \frac 1T\partial _\alpha T-u^\mu \partial _\mu
u_\alpha \right) ,  \label{161}
\end{equation}
\begin{equation}
\tau _{\alpha \beta }=\eta (\partial _\alpha u_\beta +\partial _\beta
u_\alpha -\frac 23g_{\alpha \beta }\partial _\mu u^\mu ),  \label{162}
\end{equation}
\begin{equation}
\tau =\zeta \partial _\mu u^\mu +(\zeta _1^{\prime }~,~\zeta _2^{\prime
})\partial _\mu \left( 
\begin{array}{l}
A_1^\mu  \\ 
A_2^\mu 
\end{array}
\right) ,  \label{163}
\end{equation}
where $\sigma $ is the charge-diffusion coefficient, $\zeta $ is the total
bulk viscosity coefficient, while $\kappa $ and $\eta $ are the thermal
conductivity and the shear viscosity respectively. According to the
Onsager's principle, 
\begin{equation}
\zeta _i=\zeta _i^{\prime },\qquad \qquad i=1,2  \label{164}
\end{equation}
and the positivity should be ensured by the following inequality 
\begin{equation}
\zeta (\zeta _{11}\zeta _{22}-\zeta _{12}\zeta _{21})\geqslant \zeta
_2^2\zeta _{11}-2\zeta _1\zeta _2\zeta _{12}+\zeta _1^2\zeta _{22},
\label{165}
\end{equation}
such an inequality is a generalization of that by Landau and Lifshitz \cite
{2}.

Then the full set of dissipative equations are established here, and all
transport coefficients are defined. These coefficients could be calculated
from the microscopic theory and Kubo's linear response theory could be used
here to build up the connections between the microscopic fluctuation and the
macroscopic dissipation.

In summary, the fundamental equations of motion have been constructed which
could be used to study the effects between two coupled relativistic
superfluids. The theoretical frame presented here are essentially based on
the idea of two-constituent model proposed by Landau. Furthermore, in such a
theoretical description, there arises more clearer physical significance for
the equations of motion, which relate the symmetry breaking and the physical
quantities naturally. Based on these, the interference effect and the
Josephson-type equation are discussed; the dissipation effect and entropy
production mechanism are presented, and the dissipative coefficients are
given explicitly in the first-order theory framework.

Besides the interference effect between two coupled relativistic superfluids
with spontaneous U(1) symmetry breaking, another important effect is the
discontinuity, including the shock wave propagation in the two-constituent
relativistic superfluids. Although the shock wave was studied by many
authors in the relativistic systems \cite{19}, a new theoretical frame is
still necessary to provide some new features qualitatively. Instead of
working in the frame of perfect fluid, we would rather select the
theoretical picture presented here to give a new description on the shock
wave in relativistic superfluids \cite{20}.

\end{document}